\documentclass[conference]{IEEEtranDown}
\pdfoutput=1
\IEEEoverridecommandlockouts
\usepackage{cite}
\usepackage{multicol}
\usepackage{amsmath,amssymb,amsfonts}
\usepackage{algorithmic}
\usepackage{graphicx}
\usepackage{textcomp}
\usepackage{xcolor}
\usepackage{mathtools}
\usepackage{booktabs}
\usepackage{siunitx}
\usepackage[export]{adjustbox}
\usepackage{balance}
\usepackage{subcaption}

\graphicspath{{Figures/}}

\begin{document}


\title{Performance Analysis of Cellular-V2X with Adaptive \& Selective Power Control}


\author{\IEEEauthorblockN{Md Saifuddin, Mahdi Zaman, Behrad Toghi, Yaser P. Fallah}
\IEEEauthorblockA{Connected and Autonomous Vehicle Research Lab (CAVREL)\\
University of Central Florida, Orlando, FL, USA\\
{md.saif, mahdizaman, toghi}@knights.ucf.edu, yaser.fallah@ucf.edu}

\and
\IEEEauthorblockN{Jayanthi Rao}
\IEEEauthorblockA{Research and Advanced Engineering\\
Ford Motor Company, Dearborn, MI, USA \\
jrao1@ford.com}
}

\maketitle


\begin{abstract}
 LTE based Cellular Vehicle-To-Everything (C-V2X) allows vehicles to communicate with each other directly without the need for infrastructure and is expected to be a critical enabler for connected and autonomous vehicles. V2X communication based safety applications are built on periodic broadcast of basic safety messages with vehicle state information. Vehicles use this information to identify collision threats and take appropriate countermeasures. As the vehicle density increases, these broadcasts can congest the communication channel resulting in increased packet loss; fundamentally impacting the ability to identify threats in a timely manner. To address this issue, it is important to incorporate a congestion control mechanism. Congestion management scheme based on rate and power control has proved to be effective for DSRC. In this paper, we investigate the suitability of similar congestion control to C-V2X with particular focus on transmit power control. In our evaluation, we include periodic basic safety messages and high priority event messages that are generated when an event such as hard braking occurs. Our study reveals that while power control does not improve packet delivery performance of basic safety messages, it is beneficial to high priority event message delivery. In this paper, we investigate the reasons for this behavior using simulations and analysis.
\end{abstract}

\section{Introduction}

Intelligent Transportation Systems (ITS) characterizes the next-generation mobility by fleets of connected autonomous vehicles (CAV). Large scale deployment of CAVs has yet remained on pause while their capabilities are being assessed and the research community is exploring their new advanced use cases. In order to establish policies in compliance with stringent safety requirements of urban mobility and the corresponding regulations, one of the imminent frontiers ahead is to ensure safety critical Ultra-Reliable Low Latency Communication (URLLC) service in Vehicular Ad Hoc Networks (VANET). Cellular Vehicle-to-everything (C-V2X), standardized by 3GPP is a contemporary protocol aiming to provide reliable LTE-based communication for vehicular connectivity. With increasing momentum from a large ecosystem of stakeholders consisting of auto industry, SoC manufacturers and tier-1 suppliers, it is arguably the strongest competitor of Dedicated Short Range Communication (DSRC) protocol, which has been developed in compliance with IEEE 802.11p, IEEE 1609.X and SAE J2945/1 standards \cite{sae:j2945} \cite{Fallah2018} since the reservation of 75MHz radio spectrum in 5.9GHz band for ITS services and safety applications at 1999. 

Recently FCC proposed the grant of C-V2X operations in the upper 20 MHz bandwidth of 5.905-5.925 GHz band for automotive safety applications \cite{cv2x:fcc}, possibly co-existing with DSRC. The argument on behalf of C-V2X being the main ITS protocol is further strengthened by experiment results by 5GAA \cite{cv2x:5gaa2018} and CAMP \cite{cv2x:camp2020} where LTE-based C-V2X has shown promising potential in terms of system level KPIs as latency \& reliability. Additionally, C-V2X boasts of wider network coverage, a 2-degree channel access mechanism \cite{jgozalvez:vtm}, backward compatibility and reusability of existing infrastructure \cite{wang2017comparison}, \cite{bazzi2019survey}, \cite{bazzi2017performance}.


CAVs are required to periodically broadcast their dynamic and kinematic information over the network protocol \cite{sae:j2735}. The choice of the protocol is strictly contingent upon the established latency \& reliability requirements. In vehicular context, performance of a communication protocol suffers due to the dynamic and unpredictable nature of mobility \cite{btoghi:vnc}. Specially in high density urban traffic scenario, where the number of Vehicular User Equipment (V-UE) candidates exceeds the ideal-state channel capacity, is where the safety margin of V2X protocols are pushed to the limit. During development and validation, DSRC congestion control has gone through simulation and field experimentation resulting in the algorithm as specified in SAE J2945/1. This includes modeling and analysis in a variety of simulation platforms \cite{Fallah2018} \cite{shah2019real} as well as field tests \cite{Chowdhury2018} in controlled congestion. 802.11p protocol (i.e DSRC) attempts to combat network congestion utilizing the congestion control mechanism in J2945/1 by adaptive periodic broadcasting \cite{gbansal:limericacm} \cite{clhuang:ieeenetwork} \cite{yfallah:tvtsupra2016}\cite{yfallah:mbcsyscon}. In C-V2X literature, there exists similar proposals for congestion control that utilizes beacon transmission power \cite{bkang:cv2xPowerControl} and rate adaptation \cite{ttielert:vnc}. As of now, Distributed Congestion Control (DCC) algorithm has not been specified for C-V2X. Efforts are underway in the research community to understand the suitability of J2945/1 congestion control for C-V2X \cite{aless2020congestion}. Network-level evaluation of such system is presented in \cite{mansouri2019first} and \cite{btoghi:vtc2019}, while \cite{bkang:cv2xPowerControl}, \cite{jgozalves:msn} \& \cite{jgozalvez:vtc2017spring} performed system-level analysis and suggested DCC configuration parameters for C-V2X. 

In J2945/1, congestion control consists of adaptation of the transmission power and rate as functions of vehicle density and subsequent Channel Busy Percentage (CBP). Intuitively, similar congestion control can be applicable in CV2X protocol. However, studies on the effect of transmission power adaptation on C-V2X based system is still sparse, so is a standardized congestion management scheme for C-V2X. In this paper, we analyze the consequence of transmission power adaptation on system performance using different network-level metrics which are direct expression of the reliability of network.

Beside the periodic transmission of BSMs, High Priority Messages (HPM) are transmitted in the event of sudden dynamic change of host vehicle (e.g hardbrake, lanechange); events which are of immediate concern for vehicles in vicinity. Due to their safety-critical information, HPMs have a higher priority in channel access, implied by lower latency. In this paper, congestion control is also subjected to study under coexistence of messages with different priority (here, BSM and HPM). For the remaining portion of the paper, congestion control implies adaptive power control for transmissions with a fixed broadcast rate. Our following study is divided in three main sections. In section \ref{Basic Concepts} an overview of the C-V2X system, primarily medium access (MAC) and physical layer (PHY), is provided. In section \ref{CV2X system behavior}, we discuss our previous findings about the system behavior when subjected to the proposed congestion control algorithm, and an argument has been made for the necessity of this research. Our results and corresponding analysis are presented in Section \ref{Analysis & Results} that attempts to answer the research question. We conclude with an informed opinion on the efficiency and use case of the algorithm.

\section{Basic concepts of C-V2X and congestion control} \label{Basic Concepts}
In this section, the basic significant concepts in C-V2X communication are briefly described, including the Semi-Persistent Scheduling (SPS) algorithm for resource allocation. A simple overview of congestion control algorithm in DSRC is also discussed afterwards. 

\subsection{CV2X in a nutshell}

C-V2X is a vehicular network protocol developed on 3GPP release 14 specifications. It inherits the Device-to-device (D2D) modes of communication as specified in Rel. 12 but the Medium Access Control layer (MAC) and Physical layer (PHY) are enhanced to meet the additional requirements of vehicular environment. Beside the inherited modes 1 \& 2, Rel. 14 network architecture introduces two novel modes: (i) Mode 3 when complete or partial base station (eNodeB) coverage is present, (ii) Mode 4 when no central coverage is available at mobile station i.e V-UE. The sidelink interfaces introduced in rel 14, Uu \& PC5, enables the particular V2V operations. C-V2X operates on mode 3 using Uu interface for control information and PC5 interface for databits while eNodeB distributes the channel access among the UEs in range. Outside eNodeB coverage, the operation switches to PC5 interface for the whole transaction. This mode of operation employs Semi-Persistent Scheduling (SPS) scheme to allocate radio resources among UEs in a distributed Ad-Hoc fashion. In the next subsection we provide a brief insight of the SPS algorithm. 

For understanding channel access in C-V2X, a clear depiction of the channel from PHY layer standpoint is crucial. C-V2X radio resources can be imagined as a 2D plane with time and frequency axes, where the unit granularity is 1 subframe (=1ms) and 1 subcarrier (=15kHz). Each subframe consists of 14 SC-FDMA symbols where 9 symbols are available for data transmission \cite{3gpp:36211}. Along the frequency axis, each subframe is divided into subchannels as defined in \cite{3gpp:36331}. Each subchannel is defined as a group of Resource Blocks (RB), each of which are 12 subcarrier long and $1/2$ subframe wide. RB is the tradable currency in MAC operations. Several RBs form a subchannel to accommodate a complete message block. Hence, subchannel length ($L_{subCH}$) is a function of message size and modulation and coding scheme (MCS) used in message composition. 

With HARQ scheme, C-V2X rel. 14 supports one redundant packet transmission for each data packet. A packet consists of two portion: (i) Sidelink Control Information (SCI), as format 1 standard \cite{3gpp:36212}, and (ii) Transport Block (TB), the dynamic status information. A pair of SCI \& TB can reside in adjacent or non-adjacent subchannels and in the same subframe. C-V2X adjusts the packet accordingly by employing configurable MCS and no. of subchannels for each packet. Figure \ref{fig:sensingPHY} is provided as a visual aid for understanding C-V2X PHY layer. It shows a span of $[n-(10\times P_{step}) \hspace{0.3cm} n+T_{2}]$ subframes in time with adjacent subchannelized configuration. We refer to this figure and the related parameters again in the following subsection where an overview of C-V2X MAC mechanism is provided.

\subsection{Resource selection with semi persistent scheduling (SPS)}

In this subsection, we provide a brief overview of the SPS algorithm as defined in \cite{3gpp:36321} with spotlight on the parameters helpful in understanding congestion control. A C-V2X equipped User Equipment (UE) k.pdf monitoring the channel using a moving sensing window in the past. This window is "sensed" in terms of received power associated with each resource block. When the UE has a new data packet ready for transmission and one or more transmission conditions triggers, the UE either allocates new batch of periodic resources using SPS and selects the earliest transmission opportunity, or transmits using the earliest available resource in the pool which was pre-allocated by itself using step a in a previous subframe, also known as re-using resources. Allocation of new resources is triggered by either of the following scenario: (a) available allocations do not satisfy the QoS requirement for current MAC packet data unit; (b) Certain number of consecutive transmission (Tx) opportunities remain underutilized; (c) Sidelink Resource Reselection Counter (SLRRC) reaches 0 and is reset to an integer from a predefined range. Typically, SLRRC decrements by 1 with each transmission. When it reaches 0, for the consequent transmissions, the system stochastically chooses to either perform SPS again or reclaim the resources which were allocated on previous countdown of SLRRC. In the next subframe, SLRRC is reset to $n \in N \cap [5,15]$ for 10Hz Transmission (Tx) rate.

\begin{figure}[t]
\includegraphics[width=.48\textwidth, center, trim= 5 8 8 10, clip=true]{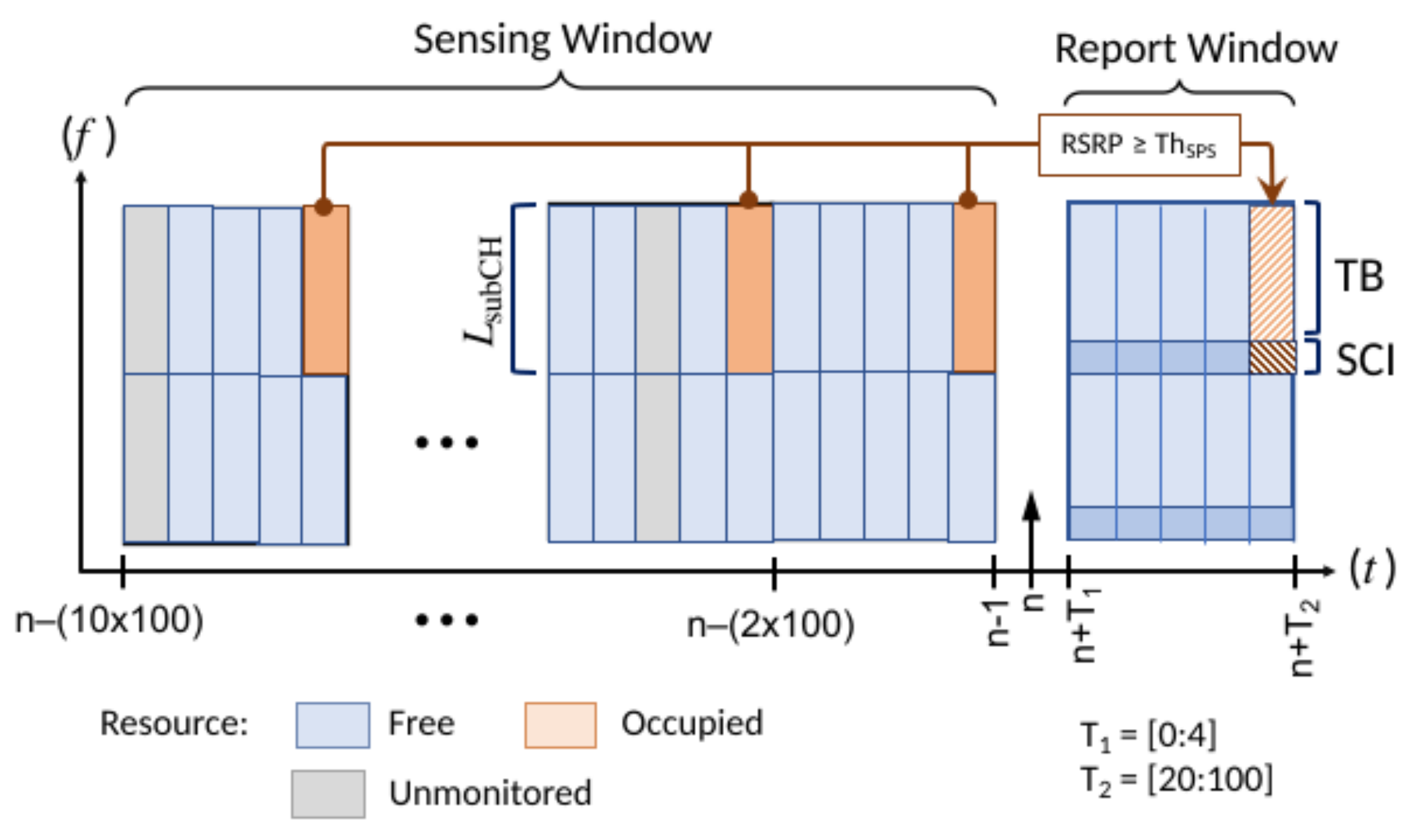}
\caption{Resource exemption in sensing procedure of SPS (SCI \& TB shown in adjacent scheme)}
\label{fig:sensingPHY}
\end{figure}

After checking the aforementioned conditions, when a UE decides to allocate new resources for its next transmission and MAC layer has received at least one packet from higher layers, it requests PHY for list of available resources. The allocation can be viewed as a feedback series of 5 s.pdf: 
\begin{itemize}
    \itemsep0em

    \item PHY considers all resources within a report window (ahead in time) as candidate resource group,
    \item Candidate resources who are periodically related with unmonitored report window resources are exempted,
    \item Resources are also exempted if they have periodic correspondence to resources in reception memory (sensing window) that are used by other UEs, confirmed with any successfully decoded SCI indicating channel occupation at those subframes. To qualify for exclusion, these resources must have Reference Signal Received Power (RSRP) over a certain threshold ($Th_{SPS}$) to be excluded.
    \item The post-exemption set of resources are reported to MAC layer. If the current set contains less than 20\% of the total resources in report window, exemption is reiterated with a 3dB increase in threshold power until the 20\% requirement is fulfilled. This increase in threshold has further implication in spatial distribution of effective network coverage as discussed in our previous work \cite{toghi:spatio}
    \item Based on Sidelink Received Signal Strength Indicator (S-RSSI), the resources are sorted in descending order (Fig \ref{fig:sensingPHY}) and MAC selects two CSRs randomly from the current resource set. These two resources are at most 15 subframes apart and selected for HARQ-0 and HARQ-1 transmission. SLRRC is reinitialized at this point and periodically corresponding resources to the selected pair are reserved for a number of future broadcasts as determined by the SLRRC.
\end{itemize}

\subsection{Congestion control algorithm}
There are ongoing efforts to develop an optimal algorithm for congestion management in CV2X. In the Wifi domain, DSRC has an established congestion control algorithm specified in J2945/1. Fundamentally congestion control is employed by:
\begin{itemize}
    \itemsep0.1em
    \item Tx Rate Control based on vehicular traffic density,
    \item Tx Power Control based on channel busy ratio (CBR)
\end{itemize}

In DSRC, the interlayer data propagation delay from MAC to PHY originates from the random back-off scheme used in CSMA-CA protocol. But it remains within degrees of microseconds. Thus rate control has a directly linear relationship with end-to-end latency, in terms of Information Age (IA). On the other hand, C-V2X MAC can add a time offset (50ms on average) in this translation due to SPS operations. However, in this paper, we keep the effect of rate control out of scope, focusing only on the effect of radiated power variation and corresponding alteration of system performance.

In case of DSRC, the power control mechanism utilizes measured CBR as a simple negative feedback control factor; by transmitting at full power when CBR is low (i.e channel is free), and at lower power when CBR is high (implying busy channel). Implementation of CBR calculation for DSRC in J2945/1, which follows 802.11p standard, is completely different than the definition of CBR in CV2X. In CV2X CBR is computed using average received power above a predefined $CBR_{Threshold}$. In terms of performance, power control scheme rewards DSRC a benefit of reduced collision boundary under lower Tx Power. Nevertheless the critical event messages, DSRC equivalent of HPM, are still serviced with higher priority in J2945/1. They are transmitted with highest Tx Power available by device at the maximum Tx rate to approach maximum information delivery. Considering the power control performance in DSRC, an investigation into CV2X using the same scheme is necessary to facilitate the search for an optimized DCC algorithm for CV2X.


\begin{figure}[b]
\vspace*{-0.3in}
\includegraphics[width=.48\textwidth, center]{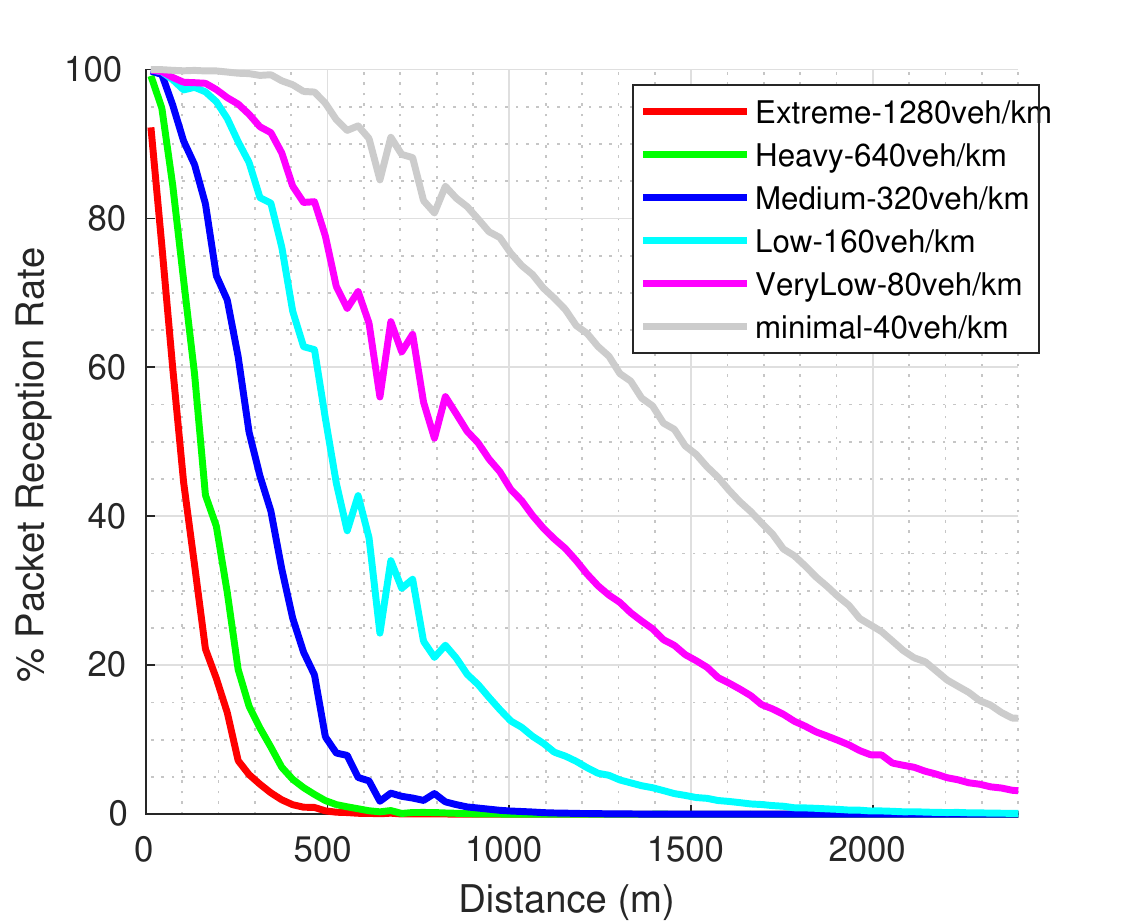}
\caption{Range of reception for different congestion scenarios in baseline (without congestion control) }
\label{fig:fig1}
\end{figure}


\section{C-V2X System Behavior with Proposed Distributed Congestion Control} \label{CV2X system behavior}
\subsection{C-V2X with DSRC J2945/1 configurations}


The proposed distributed congestion control (DCC) algorithm enables the channel access for UEs by adaptively adjusting the message transmission rate and the range of transmission. The latter is deployed by adjusting the BSM transmit power on an ad-hoc basis. 

Our previous study \cite{toghi:spatio} reveals an interesting inherent feature SPS comes with that relates the resource allocation mechanism with geo-separation of UEs present in the network at any instance. As explained in Section IIB, the PHY layer delivers a list of available CSRs to MAC, based on its selection window buffer. The final list represents the subset of CSRs carrying the least S-RSSI; accommodating idle resources as well as resources with collisions between distant UEs. Thus SPS penalizes the colliding UEs by allowing collision among geospatially sparse UE pairs, consequently realizing  maximum link budget for packet transmission for a given congestion scenario. The exclusion, sorting and selection of resources are heavily dependent upon received signal strength (RSS) at each resource unit. Thus one can expect that the channel parameters and transmission powers by individual nodes (in response to the range control mechanism) are capable of affecting the performance of the overall system. The implications are not limited to system throughput via rate control, or communication range via power control, rather they can be altering factors in channel occupancy that governs SPS.

Study \cite{btoghi:vtc2019} shows that emulating the range control mechanism employed in DSRC as standardized in J2945/1, does not generally benefit CV2X. To the best of the authors' knowledge, there is no clear study that can explain the causal relation between transmission power and the system performance of CV2X. This article attempts to explain this phenomenon in a step by step investigative manner. 

We also study and observe the impact of power control on the reception of High Priority Messages (HPM). Similar to critical event messages in J2945/1 standard, HPMs are event-driven safety critical messages which include hard braking, ABS, traction control and stability control \cite{ma_chen_refai_2009}. From link-level standpoint, HPM transmissions differ from that of BSMs in two ways: HPMs are always transmitted (1) with a tagged certificate representing the high priority and (2) at full power. A general assumption would be that HPM reception is indifferent to power adaptation since their transmission powers are not affected by it. However, the mutually shared resource pool connects them in terms of utility, hence affecting the HPM reception. This study attempts to evaluate the result of power control of the BSMs on the reception throughput of HPMs.

\subsection{Simulation Setup}

The transceiver model, channel model and receiver model used in this experiment are designed using event-based NS-3 simulator enhanced with detailed MAC \& PHY layer model in compliance with J3161/1 standard. PHY layer is modeled with resource block (RB) level abstraction using NIST proposed receiver error model. Our preliminary work \cite{btoghi:vnc} describes the details of the simulator modules. The basic configuration of parameters for the simulations in this paper are summarized in Table \ref{table:configs}.

%
\begin{table}[t]
\caption{Simulation Parameters \& Configurations}
\begin{center}
\bgroup
\def\arraystretch{1.4}
\begin{tabular*}{0.44\textwidth}{@{\extracolsep{\fill} }  l r }
\hline
\hline
Simulation Time $(T_{sim})$     & 50 second \\ 
Payload Size (with/out certificate)       & 190/300 bytes\\
MCS Index - SCI     & 2\\
MCS Index - TB (with/out certificate)          & 5 / 11\\ 
Carrier Freq.   & 5860 MHz\\     
Bandwidth & 20.00 MHz\\ 
CBR Threshold        & -92.00 dBm\\
SPS Initial Threshold & -84.18 dBm\\
Intertransmitting Time (ITT)        & 100 millisecond\\  
Propagation Loss Model & Fowlerville Model \cite{campVSC3phase1}\\

\hline
Roadlength  & 4.800 km\\
Low Density Traffic & 16.66 vehicle/km\\
Medium Density Traffic & 33.33 vehicle/km\\
Heavy Density Traffic &  66.67 vehicle/km\\
\hline
When HPM Transmissions are enabled: & \\
HPM Generating Nodes & 1\% of total\\
HPM Batch Interval & 5 second  \\
HPM Frequency  & 10 Hz \\
HPM Event Duration & 1 second \\
\hline
\end{tabular*}
\egroup
\label{table:configs}
\end{center}

\end{table}
%

\section{Analysis and Results} \label{Analysis & Results}

\subsection{C-V2X system with different transmission power}

\begin{figure*}[t]
  \begin{subfigure}{0.33\linewidth}
    \includegraphics[width=\textwidth]{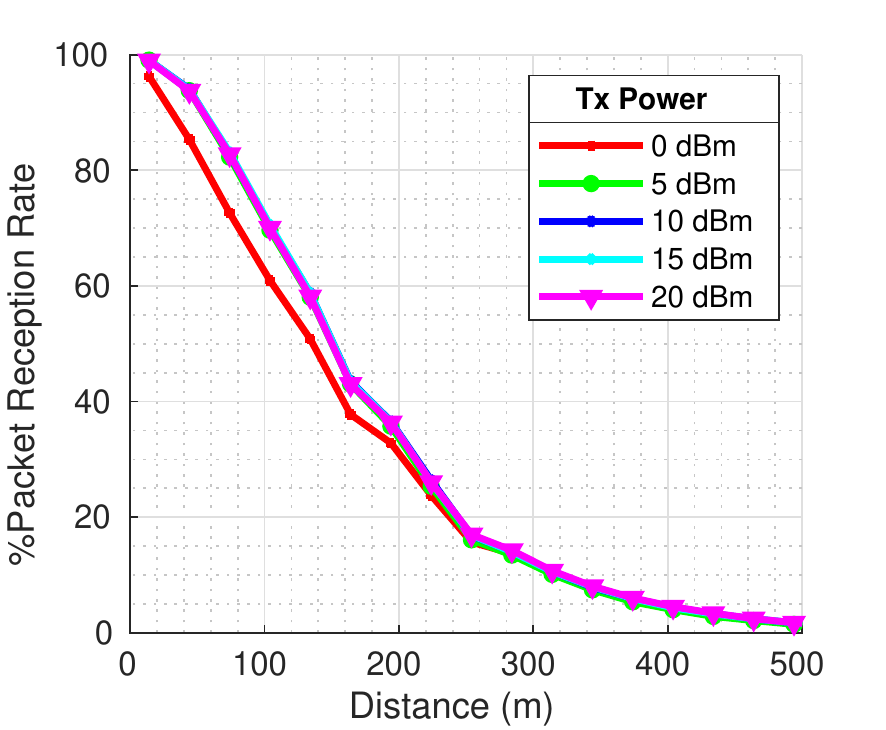}
    \caption{Heavy}
    \label{fig:prr1}
  \end{subfigure}
    %
  \begin{subfigure}{0.33\linewidth}
    \includegraphics[width=\textwidth]{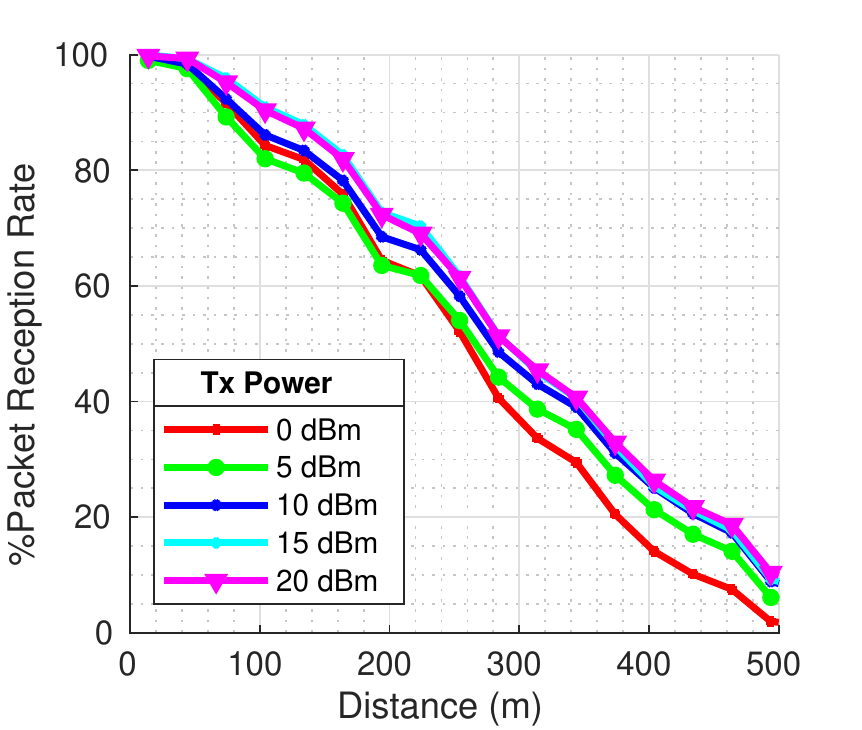}
    \caption{Medium}
    \label{fig:prr2}
  \end{subfigure}
  \begin{subfigure}{0.33\linewidth}
    \includegraphics[width=\textwidth]{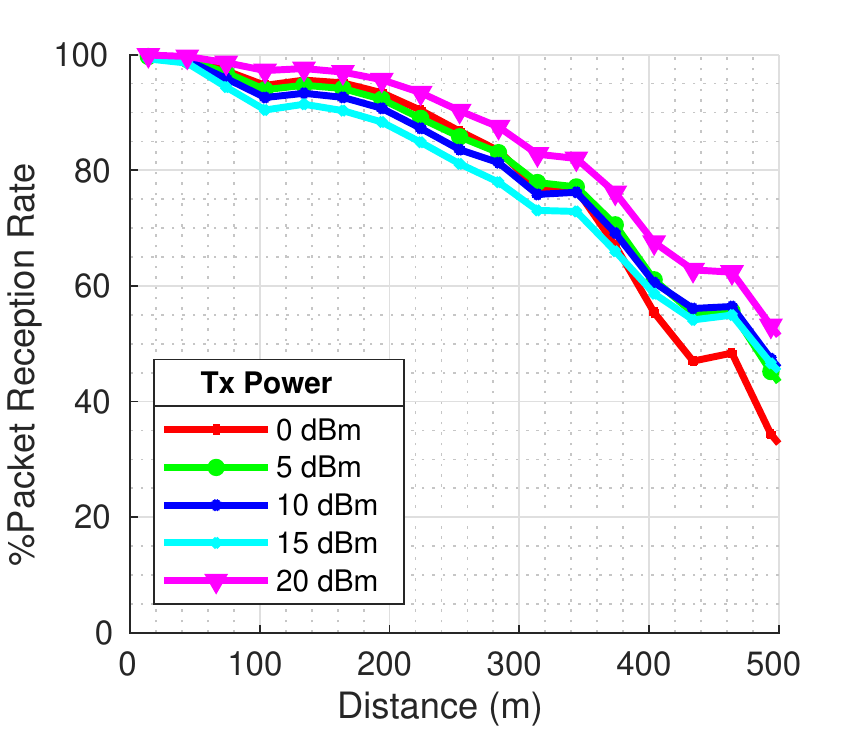}
    \caption{Low}
    \label{fig:prr3}
  \end{subfigure}
  
  \caption{Packet Reception rate for three different traffic density.} 
  \label{fig:prrfig}
\end{figure*}

To investigate the direct effect of Tx Power on the system, five different simulations has been conducted with fixed Tx Power (0, 5, 10, 15 \& 20 dBm). So all nodes in a single test broadcast with a pre-fixed power. It is important to mention that all the nodes are configured to maintain 10Hz Tx rate as per vehicular safety consensus. 

The non-line-of-sight (NLOS) and obstructed LOS radio propagation paths resulting from vehicle kinematics are modelled in our channel parameters \cite{campVSC3phase1}. Hence in our simulations, the root cause of packet loss are either SINR or collision, as \cite{jgozalvez:tvt2018} also modelled in packet reception. Figure \ref{fig:prrfig} shows the packet reception rate is Tx Power agnostic at heavy traffic density, except for very low Tx Power (0dBm). For medium and low traffic density scenario, $20dBm$ transmission results in maximum PRR, while lower Tx Power behavior varies in PRR distribution along link distances.
Higher numbers of candidate poses higher channel load and most of the packet losses are caused by resource collision. Now if Tx Power increases for all nodes, the resulting SINR under interference does not change significantly on the Rx side.  \\
Moreover, the distances of the interfering transmitters from the receivers is much lower under heavy load. So, in this case, higher Tx Power, and accordingly higher received power does not necessarily benefit the reception rate. This is further supported by 15dBm and 20dBm yielding same PRR for Medium density in figure \ref{fig:prrfig}. At low power, the reduction of interference experienced in DSRC is not present in case of CV2X due to a larger link budget and inherent range contraction during SPS resource allocation.

\begin{figure}
    \centering
    \begin{subfigure}[t]{0.26\textwidth} 
        \includegraphics[width=\textwidth, center, trim= 2 2 12 2, clip=true]{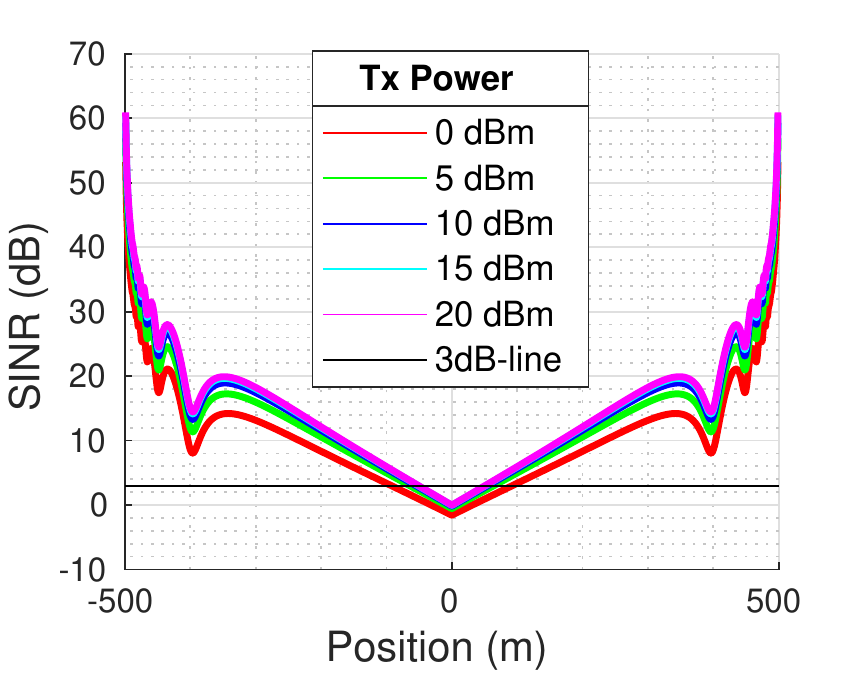}
        \caption{}
        \label{fig:SINRa}
    \end{subfigure}
   \begin{subfigure}[t]{0.22\textwidth}
        \includegraphics[width=\textwidth, center, trim= 2 2 12 0, clip=true]{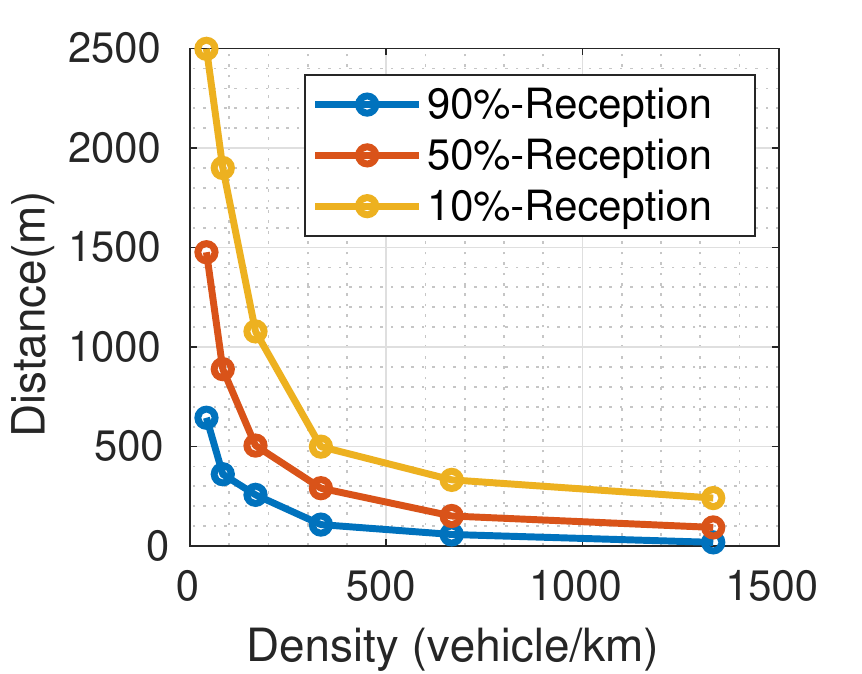}
        \caption{}
        \label{fig:SINRb}
    \end{subfigure}
    \caption{(\ref{fig:SINRa}) SINR distribution for two colliding nodes 1000m apart. (\ref{fig:SINRb}) Distance vs \%tile packet reception at different traffic densities.}
\end{figure}
%


For an algorithmic viewpoint, we need to revisit the reception mechanism of C-V2X for different Tx configuration. The SINR of a particular packet can be calculated as,
\begin{equation}
SINR = \frac{\Psi(d_0)P_{TX_0}}{\sum_{i=0}^{N_{col}}\Psi(d_i)P_{TX_i} + \eta_{NF}}
\label{SINR}
\end{equation}
Here, $\Psi(d)$ is the channel attenuation function at distance $d$, $P_{TX_i}$ is the transmission power for i-th colliding node, $\eta_{NF}$ is noise floor and $N_{col}$  number of colliding nodes. Using the given channel propagation loss from Table \ref{table:configs} channel model, we can get the average SINR for two colliding nodes at 1000m distance as plotted in Figure \ref{fig:SINRa}. The drop in SINR for $0dBm$ transmission explains the PRR drop at  $0dBm$ for heavy density plot in Figure \ref{fig:prrfig}. This also explains the maximum reception attained at $20dBm$ Tx power, which gives maximum SINR at all range bins for all densities. 

However, at lower distances ($\textless 250m$) in medium and low density traffic, the $0dBm$ Tx Power result shows slightly better PRR than some higher Tx Power scenarios. The abundance of selectable resources, after exclusion of high RSRP ( $\textgreater SPS_{Threshold}$) resources, results in such close-range gain in PRR. For very low Tx Power (not shown in the results) the system hypothetically should never reach a point, where it needs to increase the $SPS_{Threshold}$, thanks to lower Rx Power. As the Tx Power increases, the system reaches a point when the threshold needs to increase due to the 20\% resource requirement. So, there must be a Tx Power for which this requirement is marginally fulfilled. At that specific Tx Power, the SPS algorithm cannot exclude any resource by sorting based on average RSSI. 

In the following equation, x denotes the desired $SPS_{Threshold}$ at which the number of candidate-resources left becomes 20\%  of the total resources. 

\begin{equation}
\{ P_{decodeSCI}\times P_{\{RSRP_{dB} > x\}} \} = 0.8
\label{eq:SPSthres}
\end{equation}


Here, $P_{decodeSCI}$ is the decodability of one SCI on an arbitrary resource and $P_{\{RSRP_{dB} > x\}}$ is the probability of having more than $x dB$  RSRP on that resource. And in the sorting step, independent of the previous step, excess of 20\% resources with higher RSSI average falls out of selection. Tx Power 5dBm to 20dBm shows almost similar SINR profile in figure \ref{fig:SINRa}, which means $P_{decodeSCI}$ can be considered constant. As a result, equation (\ref{eq:SPSthres}) can be revised as:
\begin{equation}
P_{\{RSRP_{dB} > x\}}  \approx 0.8 /P_{decodeSCI} = constant 
\label{eq:spsthres2}
\end{equation}
\begin{figure}[h]
\includegraphics[width=.45 \textwidth, center, trim= 20 60 75 15, clip=true]{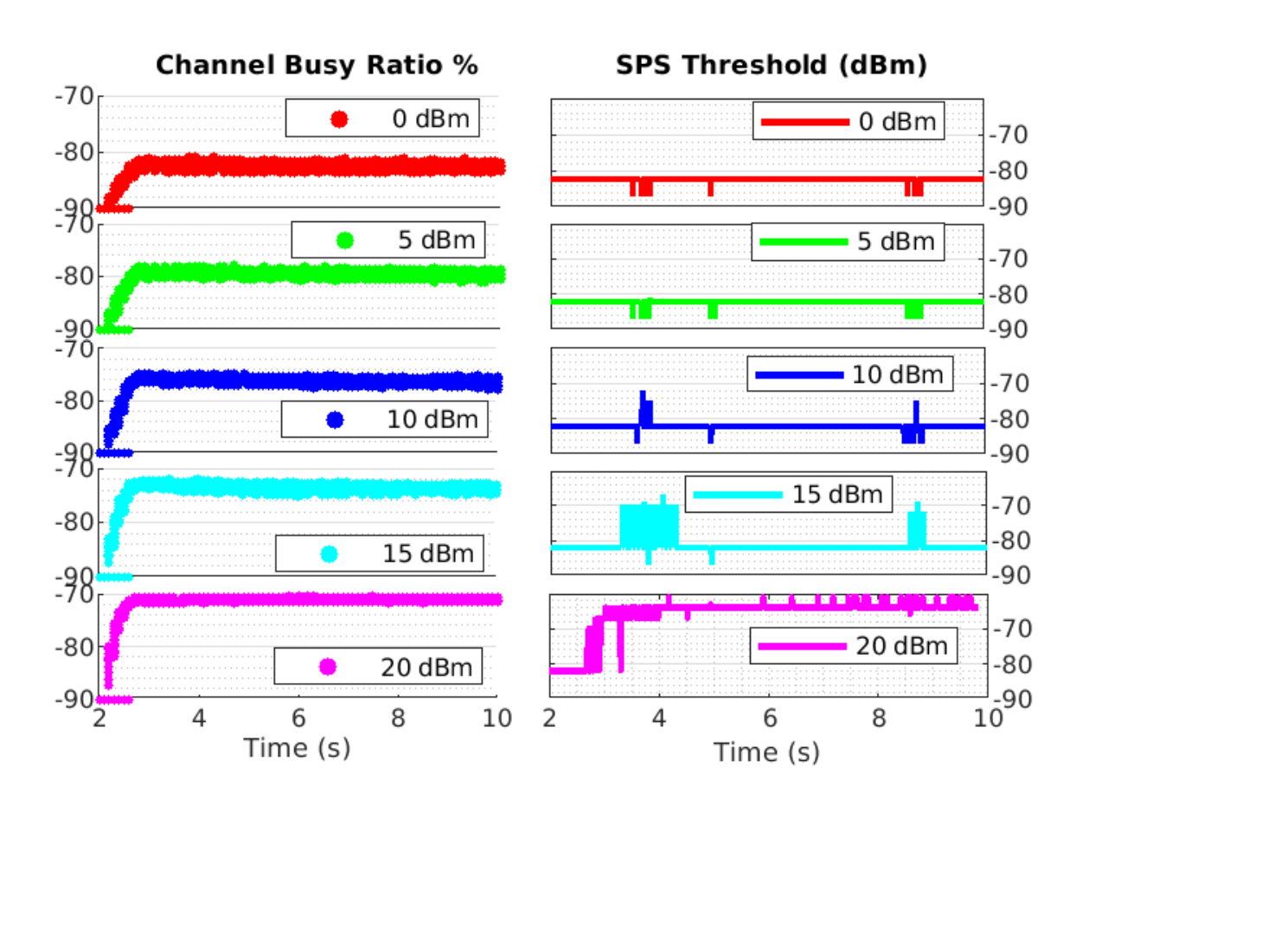}
   \caption{Low density simulation for different Tx Power (a) CBR for 50sec time (b) SPS threshold for 10sec time}
\label{fig:cbrsps}
\end{figure}

Now, observing the effective SPS threshold values for low density scenario in figure \ref{fig:cbrsps}, a correlation between measured CBR and average effective $SPS_{Threshold}$ can be established, which indirectly points to an optimum Tx Power $P_{opt}$ , at which the SPS resource retainment is exactly 20\%. Due to lack of observability, this retained resources must include a subset of all resources for which SCI decoding failed.

\begin{figure}[t]
    \includegraphics[width=.40\textwidth,  center, trim= 0 20 0 0, clip=true]{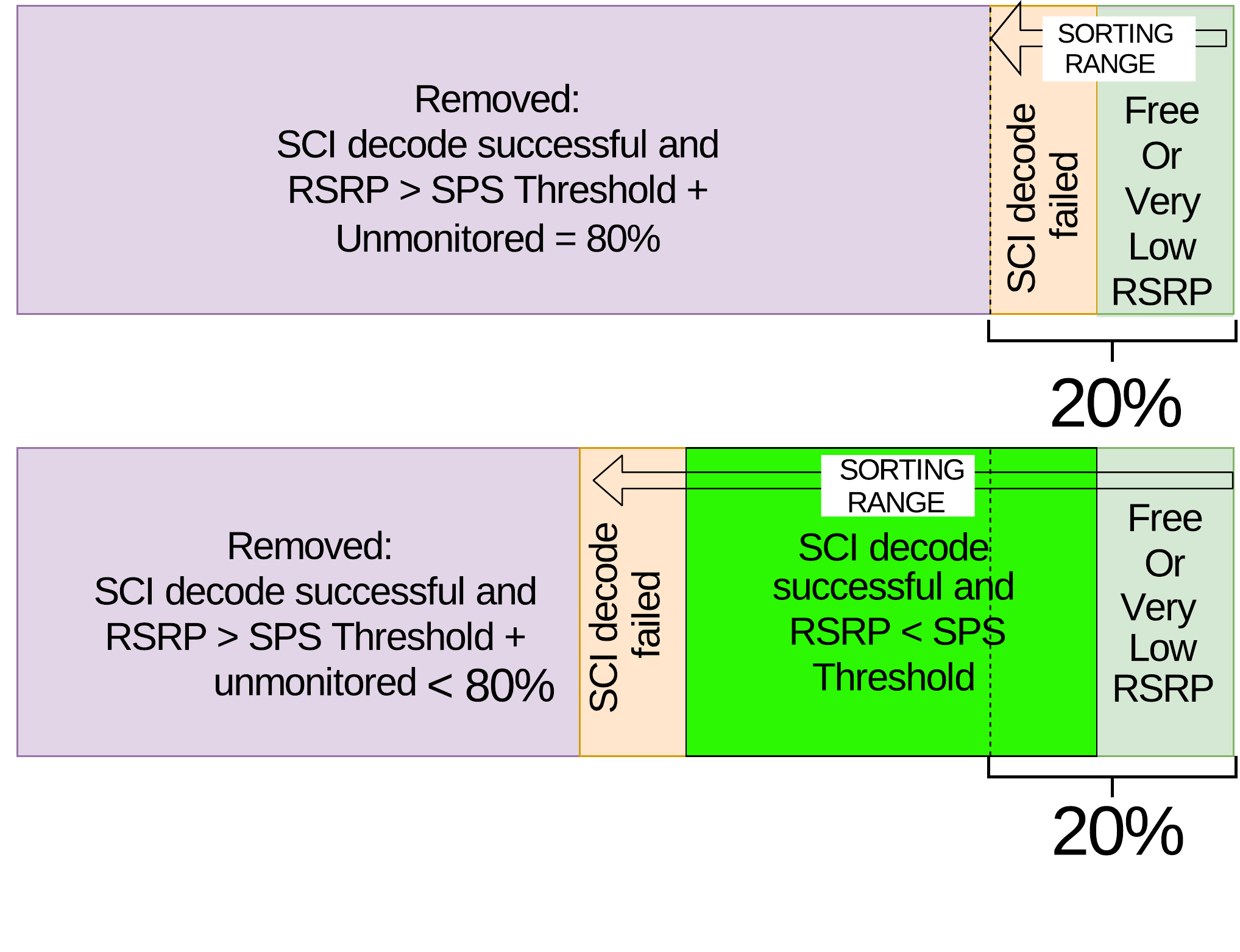}
    \caption{SPS Resource selection priorities prior to allocation, example scenario with heavy traffic (above) and low traffic density (bottom).}
    \label{fig:SPSdiag}
\end{figure}

 That means, at $P_t \approx P_{opt} $ , the benefit of sorting RSSI in the last stage of SPS loses its effect, as also shown by Manuel et al \cite{jgozalvez:tvt2018}. A snapshot of the resource table during the last selection sub-process can be seen in Figure \ref{fig:SPSdiag}. At an instance, surrounded by dense traffic population with $P_t \approx P_{opt} $ (Figure \ref{fig:SPSdiag} top), sorting does not ease the choice for the transmitting UE at all.
But, in case of  $P_t > P_{opt}$ transmit power, the SPS threshold changes according to demand of the algorithm, which changes the number of resources left before sorting based on RSSI, hence, the performance depends on the $P_t$ and induced $SINR$  instead of SPS itself in case of higher transmit powers. Similar idea was used for developing an analytical model of the SPS algorithm in \cite{jgozalvez:tvt2018}, where the SPS notion is divided into two distinct parts based on interference. 




%

\subsection{Effect of Power Control on the Higher Priority Messages }

\begin{figure}[b]
    \includegraphics[width=.48\textwidth, center, trim= 50 0 90 0, clip=true]{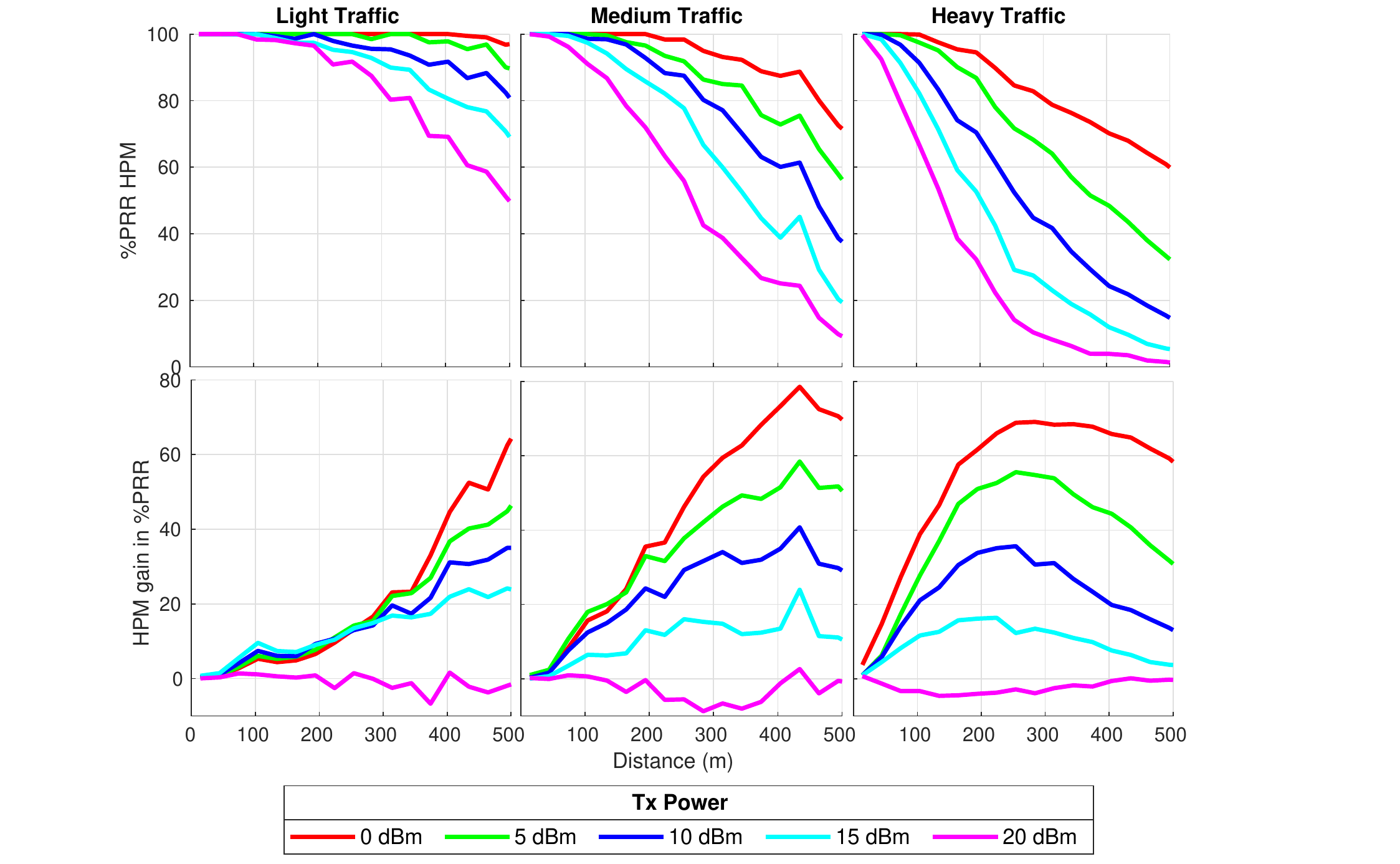}
    \caption{HPM \%PRR (top) and its improvement (bottom) over BSM \%PRR (Fig:\ref{fig:prrfig}) with different Tx Power configurations. HPM always transmits with highest Tx Power 20dBm.}
    \label{fig:diffhpm}
\end{figure}

By definition, High Priority Messages (HPM) should have higher reception probability with lower information age (IA) than regular basic safety messages. Hence they should be transmitted with the highest available rate \& Power regardless of those of regular BSM. It should also be noted that as higher priority messages are supposed to contain emergency information and so they should be aired with full certificates. Figure \ref{fig:diffhpm}  depicts that, having a higher Tx Power than the regular BSMs helps the reception of higher priority messages at all traffic condition. In light traffic, few vehicles need to deploy congestion control. So most BSMs enjoy full Tx power and thus HPMs do not get the design advantage until far range. While in comparison, UEs in medium traffic transmits more BSM under controlled power, thus more HPMs receive benefits in reception. In Figure \ref{fig:diffhpm}, results show the gain in PRR percentile HPM achieved in our simulation. The gain becomes more significant as the traffic density increases and more BSMs access the channel through DCC algorithm.     

%
%
%



Overall, we observe that adaptive adjustment of transmission power does not holistically benefit the system. However, this enables a channel headroom for high priority messages as they achieve an advantage over BSM. When BSM transmit power is adaptive, the increased HPM decoding probability achieved by their full transmit power can prove to be beneficiary for certain use cases. This holds evident for different densities.

    
    
%
\section{Concluding Remarks}
Our study establishes that transmit power control has minute impact on the performance of a Cellular-V2X network. However, high priority event messages benefit from the application of power control on BSM traffic, except in low density environments. The trade off here is the loss in packet reception for basic safety messages. Therefore, we propose to use adaptive power control only when channel access is requested by event messages. Further investigation is required to analyze possible consequences of such usage for safe deployment of cellular V2X protocol.

\balance

\bibliography{refs.bib}{}

\begin{thebibliography}{10}

\bibitem{sae:j2945}
SAE International.
\newblock On-board system requirements for v2v safety communications.
\newblock Standard Doc J2945/1, Society of Automotive Engineers, 03 2016.

\bibitem{Fallah2018}
Yaser~P. Fallah and S.~M.~Osman Gani.
\newblock Efficient and high fidelity {DSRC} simulation.
\newblock In {\em Wireless Networks}, pages 217--243. Springer International
  Publishing, October 2018.

\bibitem{cv2x:fcc}
Federal~Communications Commission.
\newblock Use of the 5.850-5.925 ghz band.
\newblock {\em FCC 19-129,
  https://docs.fcc.gov/public/attachments/DOC-360940A1.pdf}, February 2020.

\bibitem{cv2x:5gaa2018}
5g~automative association.
\newblock V2x functional and performance test report;test procedures and
  results.
\newblock {\em 5G Automative Association,
  https://5gaa.org/news/5gaa-report-shows-superior-performance-of-cellular-v2x-vs-dsrc/},
  October 2018.

\bibitem{cv2x:camp2020}
CRASH AVOIDANCE METRICS~PARTNERS LLC.
\newblock C-v2x performance assessment project.
\newblock {\em CRASH AVOIDANCE METRICS PARTNERS LLC,
  https://pronto-core-cdn.prontomarketing.com/2/wp-content/uploads/sites/2896/2020/02/CAMP\_CV2X\_SAE\_01152020\_v2.pdf},
  December 2019.

\bibitem{jgozalvez:vtm}
R.~Molina-Masegosa and J.~Gozalvez.
\newblock Lte-v for sidelink 5g v2x vehicular communications: A new 5g
  technology for short-range vehicle-to-everything communications.
\newblock {\em IEEE Vehicular Technology Magazine}, 12(4):30--39, Dec 2017.

\bibitem{wang2017comparison}
Min Wang, Martin Winbjork, Zhang Zhang, Ricardo Blasco, Hieu Do, Stefano
  Sorrentino, Marco Belleschi, and Yunpeng Zang.
\newblock Comparison of lte and dsrc-based connectivity for intelligent
  transportation systems.
\newblock In {\em 2017 IEEE 85th Vehicular Technology Conference (VTC Spring)},
  pages 1--5. IEEE, 2017.

\bibitem{bazzi2019survey}
Alessandro Bazzi, Giammarco Cecchini, Michele Menarini, Barbara~M Masini, and
  Alberto Zanella.
\newblock Survey and perspectives of vehicular wi-fi versus sidelink
  cellular-v2x in the 5g era.
\newblock {\em Future Internet}, 11(6):122, 2019.

\bibitem{bazzi2017performance}
Alessandro Bazzi, Barbara~M Masini, Alberto Zanella, and Ilaria Thibault.
\newblock On the performance of ieee 802.11 p and lte-v2v for the cooperative
  awareness of connected vehicles.
\newblock {\em IEEE Transactions on Vehicular Technology}, 66(11):10419--10432,
  2017.

\bibitem{sae:j2735}
SAE International.
\newblock Dedicated short range communications (dsrc) message set dictionary.
\newblock Standard Doc J2735, Society of Automotive Engineers, 03 2016.

\bibitem{btoghi:vnc}
B.~Toghi, M.~Saifuddin, H.~N. Mahjoub, M.~O. Mughal, Y.~P. Fallah, J.~Rao, and
  S.~Das.
\newblock Multiple access in cellular v2x: Performance analysis in highly
  congested vehicular networks.
\newblock In {\em 2018 IEEE Vehicular Networking Conference (VNC)}, pages 1--8,
  Dec 2018.

\bibitem{shah2019real}
Ghayoor Shah, Rodolfo Valiente, Nitish Gupta, SM~Osman Gani, Behrad Toghi,
  Yaser~P Fallah, and Somak~Datta Gupta.
\newblock Real-time hardware-in-the-loop emulation framework for dsrc-based
  connected vehicle applications.
\newblock In {\em 2019 IEEE 2nd Connected and Automated Vehicles Symposium
  (CAVS)}, pages 1--6. IEEE, 2019.

\bibitem{Chowdhury2018}
Mashrur Chowdhury, Mizanur Rahman, Anjan Rayamajhi, Sakib~Mahmud Khan, Mhafuzul
  Islam, Zadid Khan, and James Martin.
\newblock Lessons learned from the real-world deployment of a connected vehicle
  testbed.
\newblock {\em Transportation Research Record: Journal of the Transportation
  Research Board}, 2672(22):10--23, October 2018.

\bibitem{gbansal:limericacm}
J.~B. Kenney, G.~Bansal, and C.~E. Rohrs.
\newblock Limeric: A linear message rate control algorithm for vehicular dsrc
  systems.
\newblock {\em Proceedings of the Eighth ACM International Workshop on
  Vehicular Inter-networking}, pages 21--30, 2011.

\bibitem{clhuang:ieeenetwork}
C.~Huang, Y.~P. Fallah, R.~Sengupta, and H.~Krishnan.
\newblock Adaptive intervehicle communication control for cooperative safety
  systems.
\newblock {\em IEEE Network}, 24(1):6--13, Jan 2010.

\bibitem{yfallah:tvtsupra2016}
Y.~P. Fallah, N.~Nasiriani, and H.~Krishnan.
\newblock Stable and fair power control in vehicle safety networks.
\newblock {\em IEEE Transactions on Vehicular Technology}, 65(3):1662--1675,
  March 2016.

\bibitem{yfallah:mbcsyscon}
Y.~P. Fallah.
\newblock A model-based communication approach for distributed and connected
  vehicle safety systems.
\newblock pages 1--6, April 2016.

\bibitem{bkang:cv2xPowerControl}
B.~Kang, S.~Jung, and S.~Bahk.
\newblock Sensing-based power adaptation for cellular v2x mode 4.
\newblock In {\em 2018 IEEE International Symposium on Dynamic Spectrum Access
  Networks (DySPAN)}, pages 1--4, Oct 2018.

\bibitem{ttielert:vnc}
T.~Tielert, D.~Jiang, Q.~Chen, L.~Delgrossi, and H.~Hartenstein.
\newblock Design methodology and evaluation of rate adaptation
  bhttps://docs.fcc.gov/public/attachments/fcc-19-129a2.pdfased congestion
  control for vehicle safety communications.
\newblock pages 116--123, Nov 2011.

\bibitem{aless2020congestion}
Alessandro Bazzi.
\newblock Congestion control mechanisms in ieee 802.11p and sidelink c-v2x,
  2020.

\bibitem{mansouri2019first}
Adel Mansouri, Vincent Martinez, and J{\'e}r{\^o}me H{\"a}rri.
\newblock A first investigation of congestion control for lte-v2x mode 4.
\newblock In {\em WONS 2019, 15th IEEE Wireless On-demand Network systems and
  Services Conference}, 2019.

\bibitem{btoghi:vtc2019}
Behrad {Toghi}, Md~{Saifuddin}, Yaser~P. {Fallah}, and M.~O. {Mughal}.
\newblock {Analysis of Distributed Congestion Control in Cellular
  Vehicle-to-everything Networks}.
\newblock {\em arXiv e-prints}, page arXiv:1904.00071, Mar 2019.

\bibitem{jgozalves:msn}
R.~{Molina-Masegosa}, J.~{Gozalvez}, and M.~{Sepulcre}.
\newblock Configuration of the c-v2x mode 4 sidelink pc5 interface for
  vehicular communication.
\newblock In {\em 2018 14th International Conference on Mobile Ad-Hoc and
  Sensor Networks (MSN)}, pages 43--48, Dec 2018.

\bibitem{jgozalvez:vtc2017spring}
R.~Molina-Masegosa and J.~Gozalvez.
\newblock System level evaluation of lte-v2v mode 4 communications and its
  distributed scheduling.
\newblock In {\em 2017 IEEE 85th Vehicular Technology Conference (VTC Spring)},
  pages 1--5, June 2017.

\bibitem{3gpp:36211}
3GPP.
\newblock Physical channels and modulation (release 14).
\newblock Technical Specification (TS) 36.211, {3rd Generation Partnership
  Project (3GPP)}, 09 2018.
\newblock Version 14.8.0.

\bibitem{3gpp:36331}
3GPP.
\newblock Radio resource control (rrc); protocol specification (release 14).
\newblock Technical Specification (TS) 36.331, {3rd Generation Partnership
  Project (3GPP)}, 12 2018.
\newblock Version 14.9.0.

\bibitem{3gpp:36212}
3GPP.
\newblock Lte; evolved universal terrestrial radio access (e-utra);
  multiplexing and channel coding.
\newblock Technical Specification (TS) 36.211, {3rd Generation Partnership
  Project (3GPP)}, 04 2017.
\newblock Version 14.2.0.

\bibitem{3gpp:36321}
3GPP.
\newblock Group radio access network, evolved universal terrestrial radio
  access (e-utra), medium access control (mac) protocol specification.
\newblock Technical Specification (TS) 36.321, {3rd Generation Partnership
  Project (3GPP)}, 12 2019.
\newblock Version 15.8.0.

\bibitem{toghi:spatio}
Behrad Toghi, Md~Saifuddin, MO~Mughal, and Yaser~P Fallah.
\newblock Spatio-temporal dynamics of cellular v2x communication in dense
  vehicular networks.
\newblock In {\em 2019 IEEE 2nd Connected and Automated Vehicles Symposium
  (CAVS)}, pages 1--5. IEEE, 2019.

\bibitem{ma_chen_refai_2009}
Xiaomin Ma, Xianbo Chen, and Hazem~H Refai.
\newblock Performance and reliability of dsrc vehicular safety communication: A
  formal analysis.
\newblock {\em EURASIP Journal on Wireless Communications and Networking},
  2009(1), 2009.

\bibitem{campVSC3phase1}
CAMP~VSC3 Consortium.
\newblock Interoperability issues of vehicle-to-vehicle based safety systems
  project (v2v-interoperability) phase 1 final report.
\newblock Technical Report NHTSA-2014-0022-0029, NHTSA Publication.

\bibitem{jgozalvez:tvt2018}
M.~Gonzalez-Martin, M.~Sepulcre, R.~Molina-Masegosa, and J.~Gozalvez.
\newblock Analytical models of the performance of c-v2x mode 4 vehicular
  communications.
\newblock {\em IEEE Transactions on Vehicular Technology}, pages 1--1, 2018.

\end{thebibliography}
\bibliographystyle{unsrt}

\end{document}